# Double Beta Decay: Scintillators

**Mark C. Chen**

Department of Physics, Queen's University, Kingston, ON, K7L 3N6, Canada

mchen@queensu.ca

**Abstract**. Scintillator detectors can be used in experiments searching for neutrinoless double beta decay. A wide variety of double beta decay candidate isotopes can be made into scintillators or can be loaded into scintillators. Experimental programs developing liquid xenon, inorganic crystals, and Nd-loaded liquid scintillator are described in this review. Experiments with $^{48}$Ca and $^{150}$Nd benefit from their high endpoint which places the neutrinoless double beta decay signal above most backgrounds from natural radioactivity.

## 1. Introduction

It is widely accepted that double beta decay experiments require excellent energy resolution. The signal from neutrinoless double beta decay ($0\nu\beta\beta$) is the observation of the full energy of the decay deposited by the two emitted electrons while in ordinary double beta decay, in which two neutrinos are emitted along with the two beta electrons ($2\nu\beta\beta$), the neutrinos carry away some of the energy. In the $2\nu\beta\beta$ process the result is a continuum beta-like spectrum for the energy deposited by the electrons. Energy resolution thus helps distinguish the $0\nu\beta\beta$ energy peak at the endpoint from the $2\nu\beta\beta$ background.

The other way that energy resolution helps in double beta decay experiments is to assist in rejecting backgrounds from γ rays. A background energy spectrum will contain γ-ray lines from natural radioactivity. Common lines at known γ-ray energies are detected and identified. Lines have also been observed that were not previously expected [1]. Energy resolution is necessary in the result reported in [1] to assign some of the lines to known and unexpected backgrounds, and the line at the endpoint energy $Q_{\beta\beta}$ to the $0\nu\beta\beta$ signal.

It is interesting to consider whether a double beta decay experiment with poorer energy resolution can be competitive. The separation of $0\nu\beta\beta$ from $2\nu\beta\beta$ can exploit an endpoint spectrum fit instead of a peak search. Because the spectral shapes are well known in principle, a likelihood fit can extract limits on the presence of a $0\nu\beta\beta$ signal within a measured spectrum of $2\nu\beta\beta$ events, even if the would-be $0\nu\beta\beta$ signal is not a sharp peak. The NEMO-3 experiment makes use of this technique (as do others), as shown in Figure 1. Provided backgrounds are low and signal rates are high (i.e. sufficient statistics), the technique of fitting for the spectrum shape at the endpoint works well.

As for the need to reject background γ-ray lines using energy resolution, an experiment with poorer resolution can solve this problem by opting instead to eliminate this background. Tracking detectors, for example, aim to identify two electron tracks in order to reject γ rays that Compton scatter or γ rays that pair produce. Another way to eliminate the problem of background γ-ray lines is to look at decay energies above 2.6 MeV. This is the $^{208}$Tl line that is the highest energy γ-ray line from natural radioactivity. At energies above this value the background spectrum looks basically like a continuum and the need to sift through a forest of lines when searching for the 0νββ signal is obviated.

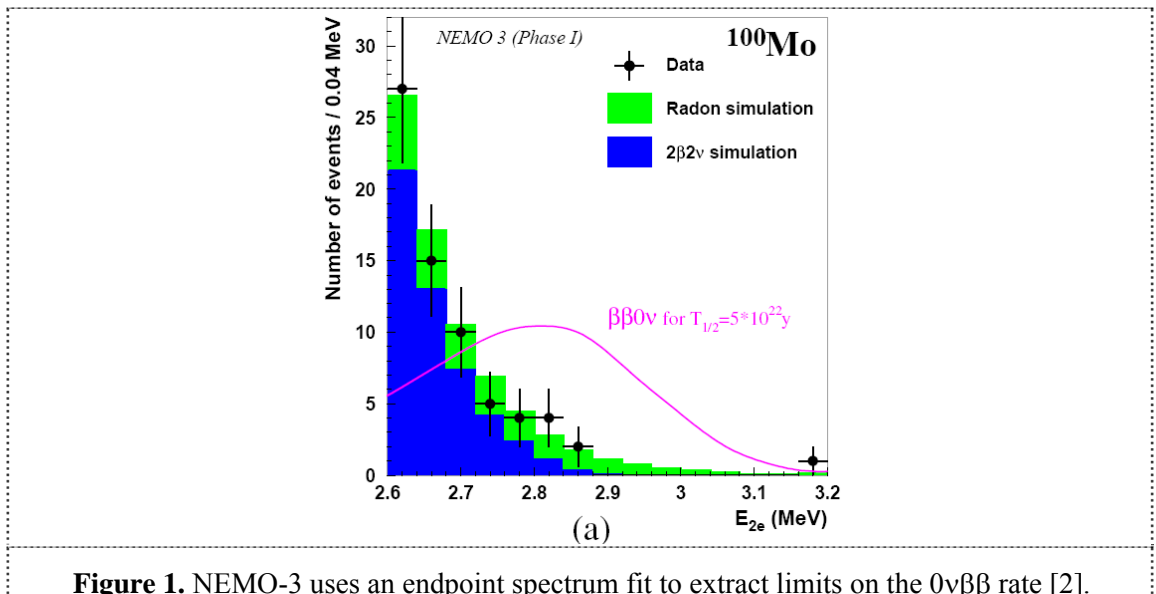

**Figure 1.** NEMO-3 uses an endpoint spectrum fit to extract limits on the 0νββ rate [2].

All of the double beta decay isotopes with $Q_{\beta\beta}$ > 2.6 MeV are listed in Table 1. The three isotopes with the highest endpoint energies have an additional advantage in that their endpoints are above the total energy deposited by β–γ backgrounds from radon. These are the double beta decay isotopes that could be utilized in a detector with poorer energy resolution without suffering from the problem of failing to distinguish between many background γ-ray lines and the 0νββ signal.

**Table 1.** Double beta decay isotopes with endpoint energies above the $^{208}$Tl line.

| isotope | $Q_{\beta\beta}$ [MeV] | natural abundance |
|---|---|---|
| $^{48}$Ca | 4.27 | 0.19% |
| $^{150}$Nd | 3.37 | 5.6% |
| $^{96}$Zr | 3.35 | 2.8% |
| $^{100}$Mo | 3.03 | 9.6% |
| $^{82}$Se | 3.00 | 9.2% |
| $^{116}$Cd | 2.80 | 7.5% |

## 2. Characteristics of scintillator detectors for double beta decay
Scintillator detectors generally have poorer energy resolution than germanium diode detectors or cryogenic bolometers. From the preceding arguments, scintillators can be considered competitive for double beta decay if they are used in experiments employing high endpoint isotopes and if backgrounds are very low. Those aspects would help a scintillator experiment cope with the poorer energy resolution. In return, the features that scintillators offer include:
- they may offer a way to fashion a large amount of an isotope into a detector economically
- several isotopes can be made into (or put in) scintillator enabling a variety of isotopes
- extremely low background environments can be achieved; phototubes stand off from the scintillator volume; self-shielding of the fiducial volume is possible; easy to model accurately
- with a liquid, there is also the possibility to purify *in-situ* to further reduce backgrounds.

In the following sections experimental double beta decay programs that use scintillators will be discussed.

### 3. Liquid xenon

The XMASS experiment is being built in the Kamioka mine. XMASS is a liquid xenon scintillation detector. In xenon, the isotope $^{136}$Xe can be used to search for double beta decay. The endpoint energy for $^{136}$Xe is 2.48 MeV. In liquid xenon there is very effective self-shielding. The research program for XMASS includes building larger and larger detectors aiming ultimately for a ~10 ton detector capable of detecting pp solar neutrinos and searching for dark matter. Double beta decay with enriched $^{136}$Xe would necessitate a separate detector because the 2νββ events would be a background to the solar neutrino and dark matter experiments.

The XMASS Collaboration is designing a version of their liquid xenon detector that is optimized for double beta decay. In their design, an elliptical tank would have liquid xenon at room temperature and high pressure at one focus of the ellipsoid and a photomultiplier tube at the other, as shown in Figure 2. This design offers excellent light collection while still allowing the PMT to be kept away from the liquid xenon fiducial volume and reducing the number of PMTs required to collect the light.

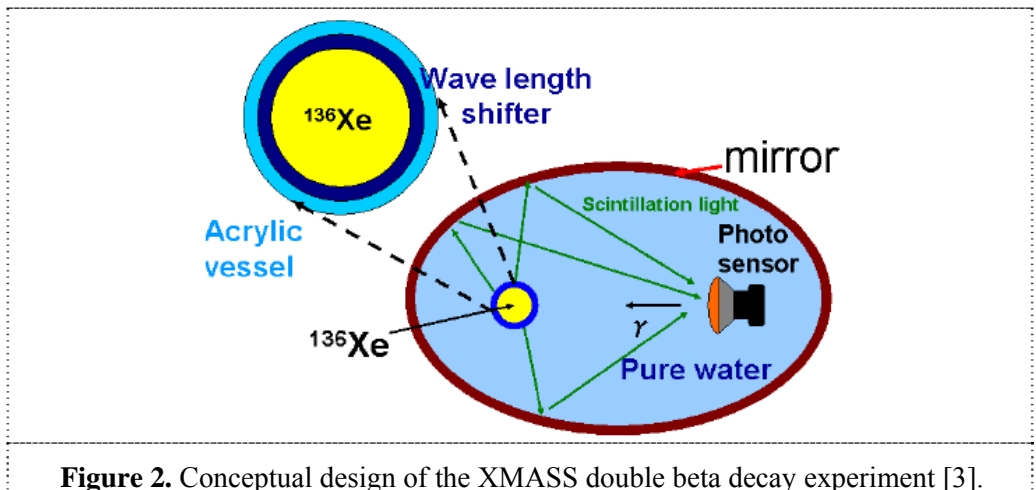

**Figure 2.** Conceptual design of the XMASS double beta decay experiment [3].

A prototype elliptical tank has been built to test this concept. Separately, the XMASS Collaboration has measured the light yield of liquid xenon at 1°C and 5.5 MPa and found it to be 64% of the light yield at the usual −100°C [3], concluding that the light yield is still very high and suitable for a double beta decay experiment.

### 4. Calcium fluoride scintillating crystals

The research group at Osaka has been using calcium fluoride scintillating crystals to search for double beta decay in $^{48}$Ca. This isotope has the highest decay endpoint but does have the drawback that the natural isotopic abundance is low. In the ELEGANT VI experiment, ~7 kg of CaF$_2$(Eu) crystals were employed but this amounted to only 7.7 g of the $^{48}$Ca isotope. Nevertheless, these experiments were successful in demonstrating that at energies above 4 MeV, backgrounds can be very low and good limits on the 0νββ half-life can be obtained.

The next step for this group is the CANDLES series of experiments. In CANDLES III the plan is to deploy ~300 kg of pure CaF$_2$ crystals (no Eu doping). In these crystals, backgrounds from the U and Th chain have been reduced; they are at the 36 μBq/kg for U and 29 μBq/kg for Th [4]. This is 30 times better than the crystals in ELEGANT VI for U and 3 times better for Th. The expected energy resolution in CANDLES III is 3.5% FWHM at $Q_{\beta\beta}$. CANDLES III will be installed in Kamioka (whereas the ELEGANT series of experiments were performed at the Oto Cosmo Observatory). The sensitivity of CANDLES III should allow exploration of Majorana effective neutrino mass down to 0.5 eV [4]. Figure 3 is a diagram of the CANDLES III detector.

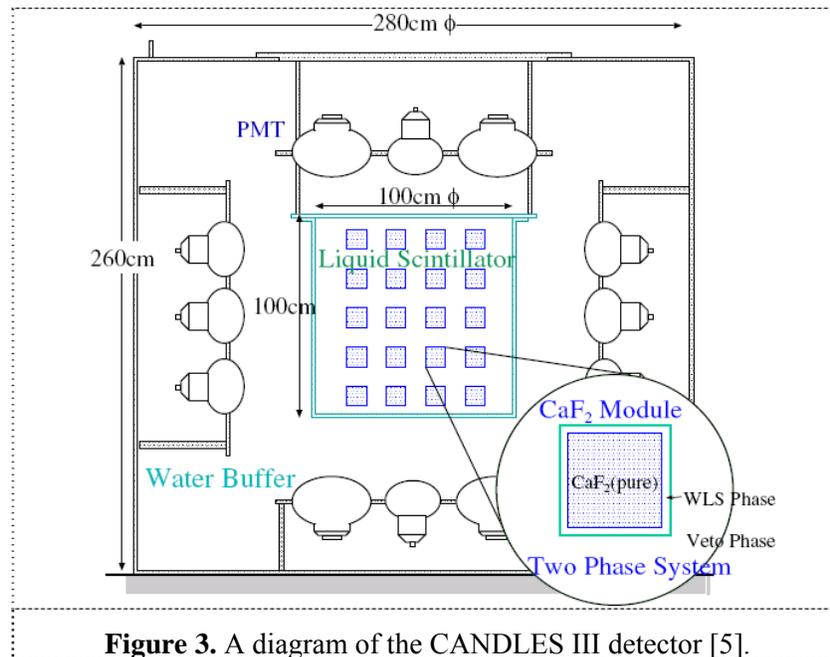

**Figure 3.** A diagram of the CANDLES III detector [5].

## 5. Inorganic crystals with Cd and Mo (Kiev group)

The Kiev group has been studying double beta decay using inorganic crystals. In the past, experiments were performed with cadmium tungstate crystals deployed in the Solotvina Underground Laboratory achieving limits on the 0νββ of $^{116}$Cd of $t_{1/2} > 1.7 \times 10^{23}$ yr (90% CL) [6]. This corresponds to a Majorana effective neutrino mass upper limit of 1.7 eV (90% CL). This group has also proposed deploying cadmium tungstate crystals in existing large water or liquid scintillator detectors, to take advantage of the low background environments available and the large photomultiplier tube coverage in existing detectors.

The Kiev group is continuing to develop scintillating crystals for double beta decay [7]. The group is continuing to improve their CdWO$_4$ crystals. Crystals as large as ~10 kg have been produced. R&D is underway to produce 1.2-1.8 kg of enriched $^{116}$CdWO$_4$ crystals in collaboration with ITEP in Moscow and the KIMS Collaboration in Korea. The group is also developing enriched $^{106}$CdWO$_4$ crystals. The isotope $^{106}$Cd is interesting because it is one of only six double positron decaying isotopes.

CaMoO$_4$ scintillating crystals have been developed by the Kiev group (the double beta decay isotope is $^{100}$Mo) [8]. The energy resolution of their best crystal is 10.3% FWHM at 662 keV [7]. A proposed experiment with enriched Ca$^{100}$MoO4 crystals is also discussed in [8]. An interesting (and somewhat regrettable) aspect of this proposed measurement is that in this search for 0νββ in $^{100}$Mo there are background contributions from the 2νββ of $^{100}$Mo *and* the 2νββ from the $^{48}$Ca in the crystal.

## 6. Nd-loaded liquid scintillator in SNO+

The Sudbury Neutrino Observatory (SNO) has concluded. The heavy water that was in the SNO detector has been removed and we plan to fill SNO with a liquid scintillator – this project is SNO+. In SNO+ we also plan to deploy a Nd-loaded liquid scintillator. 0.1% Nd loading is easily achieved and in a kiloton-sized detector this corresponds to adding an order of 1 ton of Nd. Even with natural Nd, the isotopic abundance of $^{150}$Nd is 5.6% and 1000 kg Nd thus has 56 kg of $^{150}$Nd. In comparison to

NEMO-3, which has 37 g of $^{150}$Nd, one can see that this technique provides a way to deploy a very large amount of neodymium.

The endpoint $Q_{\beta\beta}$ of $^{150}$Nd is 3.37 MeV (the second highest after $^{48}$Ca). Neodymium has the largest phase space factor of all the double beta decaying isotopes (e.g. it's a factor of 33 greater than the phase space factor for $^{76}$Ge). This implies that for the same Majorana effective neutrino mass, the 0νββ rate in $^{150}$Nd is faster. All of these advantages contribute in the desired manner to help a liquid scintillator double beta decay experiment. As discussed above, because a liquid scintillator will have poorer energy resolution, there is the need for a high $Q_{\beta\beta}$, low backgrounds (that can be achieved in a large liquid scintillator detector), and fast predicted rates (improving the statistics available in a likelihood fit of the endpoint spectrum shape). Neodymium is thus a good choice for the SNO+ experiment.

In comparison to other isotopes, it is interesting to determine how much Ge, Xe and Te would be required to obtain the same 0νββ decay rate as with 56 kg of $^{150}$Nd. Using only the phase space factor tabulated in [9] and scaling with 1/A to factor in the number of nuclei per unit mass, one calculates that 56 kg of $^{150}$Nd is equivalent to ~220 kg of $^{136}$Xe, ~230 kg of $^{130}$Te and ~950 kg of $^{76}$Ge. Including QRPA matrix elements from [10], one finds that 56 kg of $^{150}$Nd is equivalent to ~1500 kg of $^{136}$Xe (depending very much on what value is chosen for the matrix element since there is no measurement yet of the 2νββ half-life in $^{136}$Xe), ~400 kg of $^{130}$Te and ~570 kg of $^{76}$Ge. The uncertainty in all of the nuclear matrix element calculations plays a role when making such a comparison and in particular for $^{150}$Nd, which is a deformed nucleus. Nevertheless, this shows that neodymium "punches above its weight" when it comes to a neutrinoless double beta decay search.

A Monte Carlo simulation of a deployment of natural neodymium, 0.1% loading, in SNO+ is shown in Figure 4. In this simulation the predicted energy resolution at $Q_{\beta\beta}$ was 6.4% FWHM. U and Th background levels in Borexino were used in this simulation. This plot is for 3 years livetime and was simulated with a Majorana effective neutrino mass of 100 meV. Simulations of the sensitivity in SNO+ have been completed. These find a 5σ sensitivity to neutrino mass down to 100 meV with natural Nd. If we are able to obtain enriched neodymium by developing, in conjuction with our French consortium partners, an AVLIS facility in France that has the capacity to enrich hundreds of kg of material, we could improve our measurements with a larger deployment of isotope in SNO+. Our 5σ sensitivity would extend down below 40 meV with a factor of 10 enrichment.

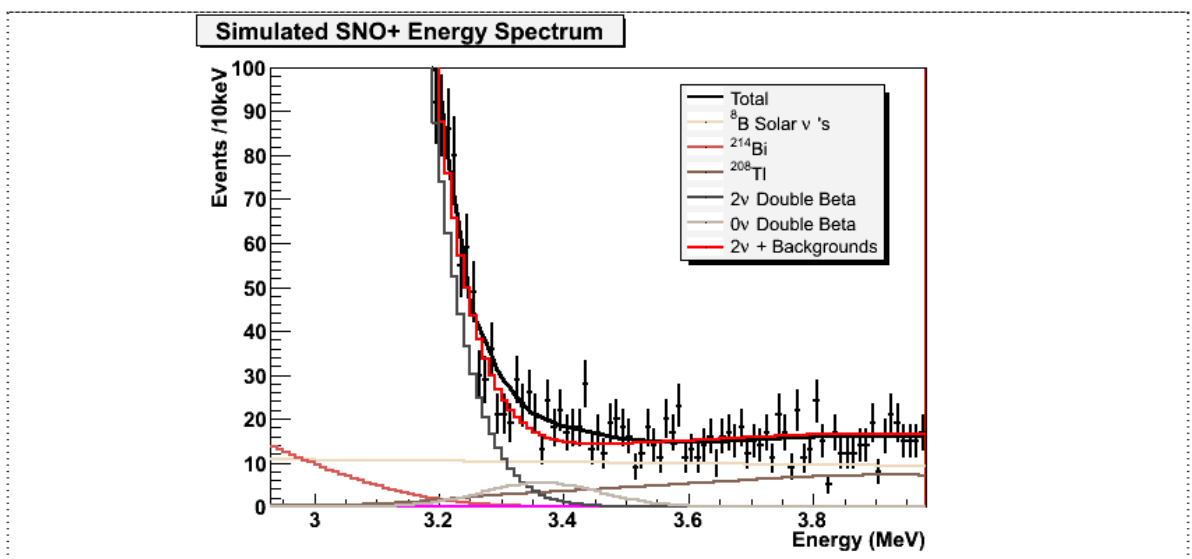

**Figure 4.** Simulated energy spectrum for SNO+ with 0.1% natural Nd. A likelihood fit extracts the contribution from (or sets limits on) the 0νββ signal. Only internal Th contamination and $^8$B solar neutrinos are significant backgrounds.

For SNO+, we have evaluated the optical properties of our Nd-loaded liquid scintillator. We have confirmed its stability; samples that have aged over one year have been compared to original optical characterization data and found to be the same. Purification techniques have been developed to remove thorium from $NdCl_3$ that is the starting material used in the synthesis of Nd-loaded liquid scintillator. These are purification techniques based upon techniques used by SNO. In SNO we added salt to the heavy water and we needed to purify the salt to remove backgrounds from U and Th down to ultra-low levels and these techniques also work for pre-purifying the neodymium salt.

The SNO+ project is preparing for construction activities in 2009, if the project receives full capital funding; SNO+ has already received partial funding for final design/engineering and initial construction. Installation of the rope net that would hold down the buoyant acrylic vessel in SNO would take place in late 2009. Construction and installation of the scintillator purification and process systems would occur in 2010. We aim for the end of 2010 for the start of scintillator filling, and soon afterward the double beta decay phase of SNO+ with Nd.

## 7. Conclusions

It is important to search for neutrinoless double beta decay with many different isotopes. Utilizing scintillators, large masses of isotopes such as $^{48}Ca$, $^{116}Cd$ and $^{150}Nd$, can be deployed. The next generation of double beta decay experiments will include several with high $Q_{\beta\beta}$ isotopes in scintillators.